\documentclass[12pt,showpacs,aps]{revtex4}
\input epsf.sty

\usepackage[sort&compress]{natbib}
\usepackage{color}

\def\hll{{\hat \ell}}
\def\ave{{\rm ave}}

\def\rr#1{\textcolor{red}{#1}}

\def\mn#1{\marginpar[\tiny{\rr{#1}}]{\tiny{\rr{#1}}}}

\def\comment#1{}
\def\aut#1{#1}

\def\mn#1{*\marginpar{*\tiny{#1}}}
\def\mn#1{}

\def\lfrac#1#2{{#1/#2}}
\usepackage{graphicx}
\begin{document}

\title{Composite Fermions and their Pair States in a Strongly-Coupled 
Fermi Liquid}
\author{Hagen Kleinert$^{(a,b)}$ and She-Sheng Xue$^{(b,c)}
\footnote{Corresponding author: xue@icra.it and shesheng.xue@gmail.com}
$}
\affiliation{$^{(a)}$Institut f{\"u}r Theoretische Physik, Freie Universit\"at Berlin, 14195 Berlin, Germany}
\affiliation{$^{(b)}$ICRANet Piazzale della Repubblica, 10 -65122, Pescara, Italy}
\affiliation{$^{(c)}$
Physics Department, University of Rome ``La Sapienza", P.le A. Moro 5, 00185 Rome, Italy}


\begin{abstract}
Our  goal is to understand the phenomena arising in optical lattice fermions  
at low temperature in an external magnetic field. 
Varying the field, the attraction between any two fermions 
can be made arbitrarily strong, 
where composite bosons form via 
so-called Feshbach resonances.
By setting up strong-coupling equations 
for fermions, we find that in spatial dimension $d>2$ 
they couple to 
bosons
which dress up fermions and lead to 
new massive composite fermions.
At low enough temperature, 
we obtain the critical temperature at which composite bosons undergo the 
Bose-Einstein condensate (BEC), leading to BEC-dressing massive fermions. 
These form tightly bound pair states which are new bosonic quasi-particles 
producing a BEC-type condensate. 
A quantum critical point is found and the formation of
condensates of complex quasi-particles is speculated over.         
\end{abstract}



\maketitle

\section{\bf Introduction}
The attraction between any two fermions 
can be tuned, 
as a function of an external magnetic field, and
be made so strong
that the coupling constant reaches the unitarity limit of infinite $s$-wave scattering length ``$a$'' via a Feshbach resonance. 
At that point, a smooth BCS-BEC crossover takes place, 
the Cooper pairs which form 
in the weak-coupling limit at low temperature
and make the system a BCS superconductor, become so strongly bound
that they behave like  
bosonic quasi-particles with a pseudogap at high temperature $T^*$, and  
form a new type of BEC 
at the critical temperature
$T_c$. 
The recent article \cite{mueller2017} reviews 
the successful progresses of BCS-BEC theories 
and experiments of 
dilute Fermi gases, whose thermodynamics can be expressed as scaling 
functions of $a$ and $T$, independently of all microscopic details. 
In this letter, as opposed to 
dilute Fermi gases, we study strongly interacting fermions in an optical lattice for ongoing experiments \cite{optical} and yet completely understood theoretical issues, such as quasi-particle spectra, phase structure and critical phenomena, as well as thermodynamical and transport properties, which can be very different from that of better-studied dilute Fermi gases. 
We use the approach of strong-coupling expansion 
to find the massive spectra of not only 
composite bosons but also composite fermions, and obtain the critical line and phase diagram in the strong-coupling region. 
Some preliminary discussions are presented on the relevance of our results to experiments.
\comment{
As the attractive coupling $g$ or the inverse scattering strength 
$1/ak_F$ (\ref{gc2tn}) increases, 
the Cooper pairs become tightly 
bound bosons. They form a normal Bose liquid, provided 
the temperature $T$ 
is less than the crossover temperature $T^*$ 
of pair formation.
Otherwise, the 
pairs dissociate into two fermions and form 
a normal Fermi liquid. 
These composite bosons undergo a BEC and become 
superfluid for $T<T_c$.
As  $1/ak_F$ increases, $T^*$ diverges away from the $T_c$. These results are summarized in 
the phase diagram of $T/\epsilon_F$ vs. $1/ak_F$\cite{ranrev2,
RZZ2012,Skleinertsc,SMRE1993}.
We study quasi-particle spectra for the phase $1/ak_F\geq 0$, and discuss the ultra-violate (UV) scaling domain 
in the unitarity limit $1/ak_F \rightarrow 0^\pm$. 
}

\vskip0.2cm
\section
{\bf Lattice fermions}
We consider fermions in an underlying lattice with a
spacing $\ell$. 
In order to address strong-coupling fermions 
at finite temperature $T$, we incorporate
the relevant $s$-wave scattering physics 
via a ``$\ell_0$-range" contact potential in the Hamiltonian for 
spinor wave function 
$\psi_{\uparrow,\downarrow}(i)$, which represents a fermionic 
neutral atom of fermion number ``$e$'' that we call ``charge'',
and 
$\psi^\dagger_{\uparrow,\downarrow}(i)$ represents its ``hole'' 
state ``$-e$'',
\begin{eqnarray}\!\!\!
\beta{\mathcal H}\! &=& \!\beta\!\!\sum_{i,\sigma=\uparrow,\downarrow}(\ell^d)\psi^\dagger_\sigma(i) \Big[-\nabla^2/(2m\ell^2) -\mu\Big]\psi_\sigma(i)\nonumber\\
\! &-& \!g \beta \sum_{i} (\ell^d)
\psi^\dagger_\uparrow(i) \psi^\dagger_\downarrow (i)\psi_\downarrow (i)\psi_\uparrow(i),
\label{Hami}
\end{eqnarray}
~\\[-1em]
$\beta=1/T$, 
each fermion field $\psi_\sigma(i)$ of length dimension $[\ell^{-d/2}]$, 
mass $m$ and chemical potential 
$\mu$ is defined at a lattice site ``$i$''. The index ``$i$'' runs over all lattice sites. The Laplace operator $\nabla^2$ is defined 
as ($\hbar=1$)
\begin{eqnarray}
\!\!\!\!\nabla^2\psi_\sigma(i)
\!&\equiv&\! \!\!\sum_{\hll}\left[\psi_\sigma(i+\hll) 
+ \psi_\sigma(i-\hll)\!-\!2 \psi_\sigma(i)\right],
\label{dalanb}
\end{eqnarray}
~\\[-1.3em]
where ${\hll}=1,\dots, d$ indicate the orientations of 
lattice spacing to the nearest neighbors.
Tuned by optical lattice and magnetic field, 
the $s$-wave attraction between the up- and down-spins is characterized by a bare coupling constant $g(\ell_0)>0$ of length dimension $[\ell^{d-1}]$ 
and the range $\ell_0<\ell$.

Inspired by strong-coupling quantum field theories \cite{Sstrong,Sxue1997}, 
we calculate 
the two-point Green functions of composite boson and fermion fields to effectively diagonalize the 
Hamiltonian into the bilinear form of these composite fields, and 
find the composite-particle spectra in the strong-coupling phase . 

\vskip0.2cm
\section
{\bf Composite bosons}\label{boson}
We first consider a composite bosonic 
field 
${\cal C}(i)=\psi_\downarrow(i)\psi_\uparrow(i)$ and 
study its two-point function on a lattice \cite{Sxue1997},
\begin{eqnarray}
G(i) = \langle \psi_\downarrow({ 0})\psi_\uparrow(({ 0}),
\psi^\dagger_\uparrow(i)\psi^\dagger_\downarrow(i)\rangle = 
\langle{\cal C}({ 0}),{\cal C}^{\dagger }(i)\rangle.
\label{bosonpc}
\end{eqnarray}
We find (methods) that in the strong-coupling effective Hamiltonian,
${\cal C}=\psi_\downarrow\psi_\uparrow$ represents
a massive composite boson 
with propagator 
\begin{eqnarray}
gG({ q})&=& {g\left[{2m/(\beta\ell)
}\right]^{2}\over 4\ell^{-2}\sum_{\hll}\sin^2(q\hll/2)+M_B^2 }
\label{scalar0}
\end{eqnarray}
with pole of mass $M_B$ and residue of form factor $gR^2_B$: 
\begin{eqnarray}
\!\!\!\!M_B^2=\left[g (2m)^2(\ell/\beta)\!-\!2d\right]\ell^{-2}>0,~
R^2_B= (2m/\beta\ell)^2. 
\label{scalar}
\end{eqnarray}  
From Eq.~(\ref{scalar0}), the effective Hamiltonian of the composite boson field
${\mathcal C}$ on a lattice can be written as
\begin{eqnarray}
\!\!\!\!\!\!\!\!{\mathcal H}^B_{\rm eff} \!&=&\!\sum_{i}(\ell^3)Z^{-1}_B{\mathcal C}^\dagger(i) \Big[\!-\!\nabla^2/(2M_B\ell^2)\! -\!\mu_B\Big]{\mathcal C}(i).
\label{eHamib}
\end{eqnarray}
The chemical potential is $\mu_B =-M_B/2$ and 
the wave function renormalization is $Z_B=gR^2_B/2M_B$. 
Provided 
$Z_B$ is finite, 
we renormalize the fermion field and the composite boson field as
\begin{eqnarray}
\psi \rightarrow (gR^2_B)^{-1/4}\psi,\quad {\rm and}\quad {\cal C} \rightarrow (2M_B)^{1/2}{\cal C},
\label{reno}
\end{eqnarray}
so that the composite boson ${\cal C}(i)$ behaves like a quasi-particle. 
This is a pair in a tightly bound state
on a lattice, analogous to the Feshbach resonance at the unitarity limit of continuum theory, and contrary to the loosely-bound 
Cooper pair 
in the weak-coupling region.

Analogously to  ${\cal C}(i)$, we consider the composite field of fermion and hole, i.e., the plasmon field ${\cal P}(i)=\psi^\dagger_\downarrow(i)\psi_\uparrow(i)$. We perform a similar
calculation to the two-point Green function 
$G_{\cal P}(i)=\langle{\cal P}(0),{\cal P}^{\dagger }(i)\rangle$,
and obtain the same result as (\ref{scalar0}) and (\ref{scalar}), indicating a tightly bound state of plasmon field, whose Hamiltonian is (\ref{eHamib}) 
with ${\mathcal C(i)}\rightarrow {\mathcal P}(i)$. 
This is not surprised since the pair field ${\cal C}(i)$ and
the plasmon field
${\cal P}(i)$ fields are symmetric in the strong-interacting Hamiltonian 
(\ref{Hami}). However, the charged pair field ${\cal C}(i)$ and neutral plasmon ${\cal P}(i)$ field can be different up to a relative phase of field $\theta(i)$. We select the relative phase field as such that $\langle|{\mathcal P}(i)|\rangle=\langle|{\mathcal C}(i)|\rangle$.
We also obtain 
the identically vanishing
two-point Green function $\langle {\cal P}(0),{\cal C}^{\dagger }(i)\rangle$, 
as ${\cal C}(i)$ is charged ($2e$) and ${\cal P}(i)$ is neutral 
$(e-e=0)$. 
 
The bound states ${\cal C}$ are composed of two constituent fermions 
$\psi_\downarrow(k_1)$ and $\psi_\uparrow(k_2)$ around the Fermi surface. Assuming $k_1\approx k_2\approx k_F$ and 
$k_2-k_1=q\ll k_F\sim \ell^{-1}$, Eq.~(\ref{scalar0}) becomes 
\begin{eqnarray}
\!\!\! g G({ q})&\!=\!& {gR_B^2/(2M_B)\over ({ q}^2/2M_B)+M_B/2 }
\Rightarrow
{{gR_B^2}\over{  q}^2+M_B^2 },~(q\ell\ll 1),
\label{scalar0_c}
\end{eqnarray}
where ``$q$'' indicates filled levels around the Fermi surface that are involved in paring.

In Eq.~(\ref{eHamib}) for $M_B^2 > 0$, 
the wave-function renormalization $Z_B=gR^2_B/2M_B\propto gT^2$ 
relates to the bound-state size $\xi_{\rm boson}$. 
As effective coupling $gT^2$ 
decreases, $Z_B$ decreases and
$\xi_{\rm boson}$ increases. The number of paring-involved states 
around Fermi surface $q\sim \xi_{\rm boson}^{-1}$ decreases.
%
The lattice pair field
${\cal C}(i)$ turns to describe
Feshbach resonance at the unitarity limit of continuum theory, 
then to describe loose Cooper pair in the weak-coupling region. 
The vanishing form factor indicates that the bosonic bound 
state pole dissolves into two fermionic constituent cut, 
i.e., two unpaired fermions,  
as discussed in Refs.~\cite{Sweinberg,doubling}. 
This qualitatively shows the feature of smooth cross-over transition from tightly bound pair field ${\mathcal C}(i)$ first 
to Feshbach resonance, then to Cooper pair, and then to unpaired 
fermions. The transition to a normal Fermi liquid of unpaired fermions takes place at 
the dissociation scale of pseudogap temperature 
$T^*(g)$ that we shall quantitatively study in future.
\comment{Limited by the validity of strong-coupling expansion, we 
were not able to calculate precisely 
dissociation temperature $T^*$ as
a function of $1/ak_F$. 
We can estimate 
at the unitarity limit $1/ak_F=0$ the crossover temperature 
$T^*\approx \epsilon_B/\log (\epsilon_B/\epsilon_F)^{3/2}$ 
\cite{SMRE1993}.  
The binding energy $\epsilon_B/\epsilon_F=2f_-/f_+, f_\pm = \sqrt{1+\hat\mu^2}\pm\hat\mu$ from Eq.~(3.285)  
of textbook \cite{SCQF}. Here we insert for the crossover
parameter $\hat\mu=\mu_B/M_B=-1/2$, using $M_B$ as 
the mass gap at $1/ak_F=0$ and find 
$\epsilon_B/\epsilon_F\approx 5.24$ and $T^*/\epsilon_F\approx 4.86$.}

\vskip0.2cm
\section
{\bf Phase transition}
On the other hand, as the effective coupling $gT^2$ varies, 
the mass term $M_B^2{\cal C}{\cal C}^\dagger$ 
in Eqs.~(\ref{scalar0}) and (\ref{scalar}) possibly changes its sign
from $M_B^2>0$ to $M_B^2<0$ and the pole $M_B$ becomes imaginary,  implying 
the second-order phase transition from the symmetric phase to the condensate phase \cite{Sxue1997}, where the nonzero condensation 
of pairing field $\langle {\mathcal C(i)}\rangle\not=0$ is developed and will be duly discussed below. 
$M_B^2=0$ gives 
the critical line: 
\begin{eqnarray}
m^2g_cT_c= d/ (2\ell),
\label{criticgt}
\end{eqnarray} 
where $g_c$ is the critical value of bare coupling $g(\ell_0)$ 
defined at the short-distance scale $\ell_0<\ell$, 
and $T_c$ is the critical temperature. 
Note that this result is qualitatively consistent with the 
strong-coupling behavior $T_c \sim t^2/U$ in the attractive 
Hubbard model \cite{MRR1990}.

In order to discuss the critical behaviors of the second-order 
phase transition, we focus our attention on the neighborhood 
(scaling domain) of the critical line (\ref{criticgt}), where 
the characteristic correlation length $\xi\sim M_B^{-1}$ 
is much larger than the lattice spacing $\ell$, thus 
microscopic details of lattice are physically irrelevant. 
Therefore, in the scaling domain of 
critical line (\ref{criticgt}) we approximately treat the lattice field theory (\ref{Hami}) and (\ref{dalanb}) as a continuum field theory 
describing Grand Canonical Ensembles at finite temperature \cite{Skleinertsc, pion}. In this framework, we obtain the 
relation between the critical ``bare'' coupling $g_c$ 
in Eq.~(\ref{criticgt}) and the ``renormalized'' coupling described by 
the $s$-wave scattering length $a$ via the two-particle Schr\"odinger equation at critical temperature $T_c$, 
\begin{eqnarray}\!\!\!\!\!\!\!\!\!\!\!\!\!\!\!\!\!\!\!\!\!\!
~~~~\frac{m}{4\pi a} \!=\! -\frac{1}{g_c(\Lambda)}+\frac{T_c}{V}
\sum_{\omega_n,|{\bf k}|<\Lambda}
\frac{1}{\omega_n^2+\epsilon^2_{\bf k}},~~\Lambda=\pi \ell_0^{-1}.
\comment{
=-\frac{1}{g(\Lambda)}+
\frac{1}{V}
\sum_{|{\bf k}|<\Lambda}
\frac{1}{2\epsilon_{\bf k}}\tanh\frac{\epsilon_{\bf k}}{2T}.}
\label{gc2t}
\end{eqnarray}
Here continuum spectrum $\epsilon_{\bf k}=|{\bf k}|^2/2m$ denotes the
energy of free fermions and $\sum_{\omega_n,|{\bf k}|<\Lambda}$ contains 
the phase-space integral and the sum over the {\it Matsubara frequencies} 
$\omega_n=2\pi T n$, 
obtained
in Eq.~(7A.3)
of  
textbook \cite{SCQF}.
This can be written
\comment{for $d=3$ as
$\frac{4\pi}{(2\pi\hbar)^3}\int d|k| m \tanh \frac{|k|^2}{4mT}$. 
}%
in $d$ dimensions as
{
\begin{eqnarray}\!\!\!\!\!\!\!\!\!\!\!\!\!\!\!\!\!\!\!\!\!\!
~~~~~~~~~~~
\!\!\!\!\!\!\!\!\!\!\!\!\!
\frac{4\pi a}{mg_c(\Lambda)}
&=& \frac{ak_F}{4\pi b
}{\mathcal S}_d(T_c)-1.
\label{gc2tn}
\end{eqnarray}
\comment{where   $\frac{ak_F}{4\pi b d \hbar^3}
{\mathcal S}_d(T)= \frac{4\pi a}{mV}
\sum_{|{\bf k}|<\Lambda}
\frac{1}{2\epsilon_{\bf k}}\tanh\frac{\epsilon_{\bf k}}{2T}$
is the temperature-dependent phase-space
sum in (\ref{gc2t}).}
In $d=3$ dimensions, we approximately adopt the 
half-filling fermion density 
$n\approx 1/\ell^3\approx k_F^3/3\pi^2
$, Fermi 
momentum $k_F\approx (3\pi^2
)^{1/3}/\ell$, and Fermi
energy $\epsilon_F=k_F^2/2m$.
Moreover we introduce the dimensionless and optical lattice tunable  
parameter $b= 2^{-1}(3\pi^2
)^{1/3} \ell_0/\ell < 1$, that measures
the fraction of filled levels around $\epsilon_F$, 
contributing to the pairing, and find 
${\mathcal S}_3(T_c)\equiv  \int_0^1dt \tanh\left[\frac{\epsilon_F}
{T_c}\frac{\pi^2 t^2}{8b^2}\right]$, with  $S_3(0)\!=\!1$. 
Equations (\ref{gc2t}) and (\ref{gc2tn}) reduce properly to the well-known $a-g(\Lambda)$ relation in the continuum theory at zero temperature. Nevertheless, it is worthwhile in future study to use the lattice version of Eq.~(\ref{gc2t}), where the continuum spectrum 
$\epsilon_{\bf k}$ is replaced by the lattice spectrum (\ref{dalanb}) 
of free fermions and the phase-space summation 
$\sum_{\omega_n,|{\bf k}|}$ is over entire Brillouin zone. We expect 
a numerical modification of the phase-space function 
${\mathcal S}_d(T_c)$ in Eq.~(\ref{gc2tn}), and it does not 
qualitatively change the critical behavior discussed below, since the critical line (\ref{criticgt}) is obtained from lattice calculations.   

From the critical line (\ref{criticgt}) 
and approximate $a-g_c(\Lambda)$ relation (\ref{gc2tn}) 
at critical temperature $T_c$, 
we find for large $1/ak_F$ or $g_c(\Lambda)$,
\begin{equation}
T_c=T^u_c(T_c)
\left[1-\frac{4\pi 
b}{ {\mathcal  S}_d(T_c)} 
\frac{1}{ak_F}\right],
\label{26h}
\end{equation}
where $T^u_c(T_c)/\epsilon_F\equiv 
(3\pi^2)^{-1/d} \lfrac{d {\mathcal S}_d(T_c)}{
(4\pi)^2 
b}$ and $T^u_c=T^u_c(T^u_c)$ is the critical temperature at $1/ak_F=0$.
In the superfluid phase $(T<T_c)$,
the boson field ${\mathcal C}(i)$
develops a nonzero 
expectation value $\langle{\mathcal C}(i)\rangle$ and undergoes 
BEC. To illustrate the critical line (\ref{26h}) 
separating two phases, in Fig.~\ref{f:newphasediagram}, we plot numerical results of (\ref{26h}) in $d=3$ for the parameters 
$b=0.02, 0.03$ corresponding to the ratios 
$\ell_0/\ell=0.013,0.02$. The `` linear'' critical line
in Fig.~\ref{f:newphasediagram} 
is due to the weak $T_c$-dependence in the 
highly nonlinear relation (\ref{26h}).

Our strong-coupling result shows 
a decreasing critical temperature $T_c$ for large $g_c$ or $1/ak_F$, 
where the length $a$ and size $\xi_{\rm boson}$ can approach the lattice 
spacing $\ell$ and become even smaller.
Taking the limit  
$g_c\rightarrow\infty$ at constant $T_cg_c$, we find
$T_c\rightarrow 0$, implying a {\it quantum critical point}.
This happens at
$1/ak_F\rightarrow (1/ak_F)_{\rm qc}\equiv {\mathcal S}_d(0)/4\pi b
=1/4\pi b
$, where composite particles are most tightly bound states, locating at the lowest energy level of the ``$\ell_0$-range'' contact potential with $a=2\pi \ell_0$, their thermal fluctuations are negligible. 
The figure shows that for smaller $b$, more fermions in the
Fermi sphere are involved in the paring, resulting in a larger
composite-boson density and a higher $T_c$.


Our results (Fig.~\ref{f:newphasediagram}) 
are valid only in the very strong-coupling region 
$g_c \gg 1$, i.e., $0< 1/ak_F < (1/ak_F)_{\rm qc}$, 
since our approach of strong-coupling expansion to interacting 
lattice fermions is particularly appropriate in this region. 
Nevertheless, we extrapolate the critical line in Fig.~\ref{f:newphasediagram} to the critical temperature $T^u_c/\epsilon_F\approx 0.31, 0.2$ at the unitarity limit $1/ak_F=0$, so as to compare with 
and contrast to the results obtained by continuum theory for dilute Fermi gas. 
The experimental value 
$T^u_c/\epsilon_F\approx 0.167$ of dilute Fermi gas \cite{martin2012} can be achieved for $b\approx 0.037$. However, such a simple 
extrapolation is not expected to be quantitatively correct, 
and high-order strong-coupling calculations 
(${\mathcal O}(1/g^n), n>1$) and proper summation over `` $n$'' 
are needed, since the critical coupling $g_c$ is not very large in the neighborhood of the unitarity limit $1/ak_F=0$. 
In fact, for $1/ak_F\gtrsim 0$ our result of monotonically 
decreasing $T_c/\epsilon_F$, as shown in Fig.~\ref{f:newphasediagram}, 
is in contrast not only with the constancy BEC limit 
$T_c/\epsilon_F=0.218$ obtained by considering mean-field 
value and Gaussian fluctuation 
in the normal states of Fermi gas \cite{SMRE1993}, 
but also with non-monotonic $T_c/\epsilon_F$ obtained 
by Quantum Monte Carlo simulations \cite{QMC}.

However, we have to stress the validity of our results of 
monotonically decreasing $T_c/\epsilon_F$ that leads to a 
quantum critical point in very strong-coupling region 
($0< 1/ak_F < (1/ak_F)_{\rm qc}$), i.e., 
beyond the unitarity limit $1/ak_F=0$.
The reasons are that these results are obtained by considering 
the strong-coupling limit 
$(g\rightarrow\infty)$ of interacting lattice fermions, calculating the first order correction ${\mathcal O}(1/g)$ and nontrivial recursion relation (see methods). Our calculations take into account contributions from strong-correlating collective modes of lattice fermions at the strong-coupling limit. 
Therefore, our approach is very different from the approach of
considering Gaussian fluctuations of normal states of continuum fermion gas upon the mean-field limit, which is not valid in the 
strong-coupling limit. It is deserved to use Quantum Monte Carlo simulations to study the lattice field theory
(\ref{Hami}) and (\ref{dalanb}) in such a very strong-coupling region.

To end this section, we make the following speculations 
on the phases beyond 
the quantum critical point $(1/ak_F)_{\rm qc}$. 
The quantum phase transition undergoes from the phase 
$(1/ak_F)<(1/ak_F)_{\rm qc}$ to the phase $(1/ak_F)>(1/ak_F)_{\rm qc}$, 
and the latter possibly involves the formation and condensation of more complex 
composite quasi-particles of higher spin-angular momentum pairing, 
e.g., spin triplet ${\mathcal C}^{\rm tri}\equiv (\psi_\uparrow\psi_\uparrow, \psi_\uparrow\psi_\downarrow, \psi_\downarrow\psi_\downarrow)$
with mass gap $M_B^{\rm tri}(T)$ and phase factor ${\mathcal S}_d^{\rm tri}(T)$, 
and the bosonic triplet ${\mathcal C}^{\rm tri}$ dresses up a 
fermion to form a three-fermion state discussed below.
The critical line $T^{\rm tri}_c(1/ak_F)$ 
obtained from $M_B^{\rm tri}(T_c^{\rm tri})=0$  
separates the quasi-particle ${\mathcal C}^{\rm tri}$ formation from their condensate phases. This critical line $T^{\rm tri}_c(1/ak_F)$
starts from the quantum critical point $T^{\rm tri}_c=0$ and 
$(1/ak_F)=(1/ak_F)_{\rm qc}$, and increases as $(1/ak_F)$ 
increases ($g$ decreases) for $(1/ak_F)>(1/ak_F)_{qc}$. 
The first-order phase transition should occur at the quantum critical point, which should be an IR fixed point. All these aspects will be studied in future.

\begin{figure}   
\begin{center}
\vspace{-3em}
\includegraphics[width=0.55
\textwidth]{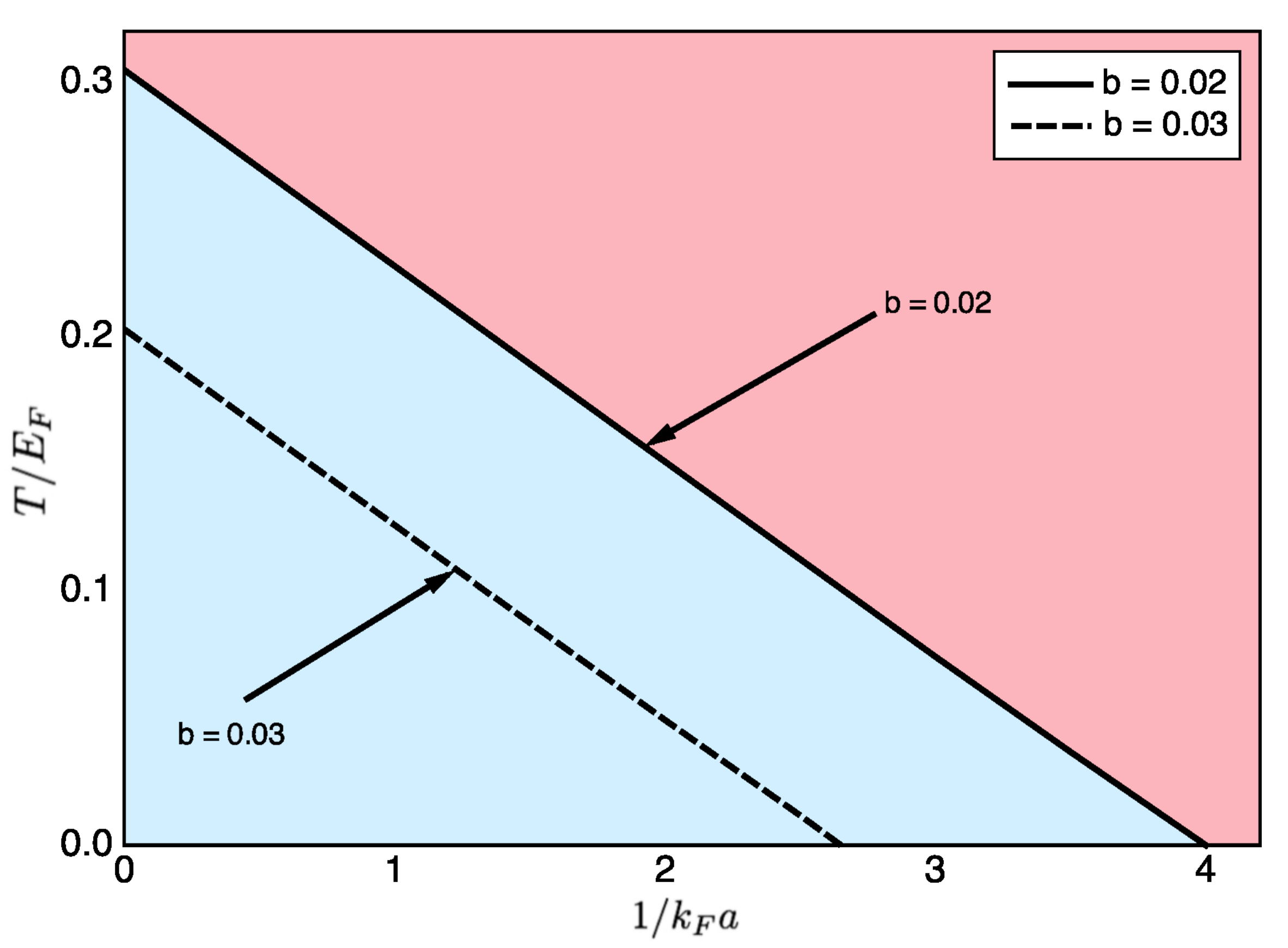}
\vspace{-1em}
\caption{
{\bf The phase diagram in strong coupling region.} 
Transition temperature 
$T_c/\epsilon_F$ is plotted
as a function of
$1/k_F a\ge 0$ 
for the 
selected 
parameters
 $b=0.02, 0.03$.
The  quantum  critical point is shown to lie at
one of the corresponding zeros
 $(1/k_F a)_{T_c=0}=4.0,2.7$. 
Above the critical line is
a normal liquid consisting of
massive composite bosons
and fermions. Below the critical line lies
a superfluid phase with a new type 
of BEC involving composite massive fermions.
}\label{f:newphasediagram}
\end{center}
\vspace{-2em}
\end{figure}

\comment{ e.g., spin triplet ${\mathcal C}^{\rm tri}\equiv (\psi_\uparrow\psi_\uparrow, \psi_\uparrow\psi_\downarrow, \psi_\downarrow\psi_\downarrow)$ with mass gap $M_B^{\rm tri}(T)$ and phase factor ${\mathcal S}_d^{\rm tri}(T)$, 
which are {\it not} given by Eqs.~(\ref{gc2t}) and (\ref{scalar}), 
and the bosonic triplet ${\mathcal C}^{\rm tri}$ dresses up a 
fermion to form a three-fermion state discussed below.
The critical line $T^{\rm tri}(1/ak_F)$ from $M_B^{\rm tri}(T)=0$  
separates the quasi-particle ${\mathcal C}^{\rm tri}$ formation from the condensate phases. Starting from the quantum critical point 
it increases as $T$ and $(1/ak_F)$ increase ($g$ decreases) 
for $(1/ak_F)>(1/ak_F)_{qc}$.  
Viewing the four-fermion 
interaction as an attractive potential, this ``infinite'' coupling point indicates
the most tightly bound state locating at the lowest energy level 
of the potential, with a scattering length $a=2\pi \ell_0$. 
If the attraction comes from a $\delta$
-function, the length parameters
$a$, and $b$ vanish, while
 $(1/k_Fa)_{qc}\rightarrow \infty$, recovering the nearly
horizontal critical line presented in Figure 3 in Ref.~\cite{ranrev2}. 
}

\vskip0.2cm
\section
{\bf Composite Fermions}
\hskip0.01cm
In this section, we show that the strong-coupling attraction between fermions forms not only composite bosons but also composite fermions. 
To exhibit the presence of composite fermions, 
using
the paring field 
${\mathcal C}(i)$ 
we calculate the two-point Green functions \cite{Sxue1997}:
\begin{eqnarray}
\hspace{-3em}
S_{LL}(i)\hspace{-3pt}&\equiv&\hspace{-3pt}\langle\psi_\uparrow(0),\psi^\dagger_\uparrow(i)\rangle,\label{csll}\\
\hspace{-3em}
S_{ML}(i)\hspace{-3pt}&\equiv&\hspace{-3pt}
\langle \psi_\uparrow(0) ,{\mathcal C}^\dagger(i)\psi_\downarrow(i)\rangle 
,
\label{csml}\\
\hspace{-3em}S^\dagger_{ML}(i)\hspace{-3pt}&\equiv&\hspace{-3pt}
\langle\psi^\dagger_\downarrow(0) {\mathcal C}(0), 
\psi^\dagger_\uparrow(i)\rangle 
\label{csml+}\\
\hspace{-3em}S_{MM}(i)\hspace{-3pt}&\equiv&\hspace{-3pt}
\langle \psi^\dagger_\downarrow(0) {\mathcal C}(0) ,{\mathcal C}^\dagger(i)\psi_\downarrow(i)\rangle. 
\label{csmm}
\end{eqnarray}  
We find (methods) that in the strong-coupling effective Hamiltonian,
the propagator
\begin{eqnarray}
S_{\rm Fermion}(p)&=&{2\over 4\ell^{-2}\sum_{\hll}\sin^2(p\hll/2)+M_F^2 },
\label{cpropf}
\end{eqnarray}
represents a composite 
fermion that is the superposition of the fermion $\psi_\uparrow$ and the three-fermion state ${\mathcal C}(i)\psi^\dagger_\downarrow(i)$ \cite{Sxue1997},
\begin{eqnarray}
\Psi_\uparrow(i)&=&R_B^{-1/2}\psi_\uparrow(i) + R_B^{-3/2}{\mathcal C}(i)\psi^\dagger_\downarrow(i)\nonumber\\
&\Rightarrow & g^{1/4}\psi_\uparrow(i) 
+ g^{3/4}{\mathcal C}(i)\psi^\dagger_\downarrow(i),
\label{ccomf}
\end{eqnarray}
where the three-fermion state ${\mathcal C}(i)\psi^\dagger_\downarrow(i)$ is made of a hole $\psi^\dagger_\downarrow(i)$ ``dressed'' by a cloud of composite bosons ${\mathcal C}(i)$.
The associated  two-point Green function reads
\begin{eqnarray}
&&\langle \Psi_\uparrow(0) ,\Psi^\dagger_\uparrow(i)\rangle
=\langle \psi_\uparrow(0), \psi^\dagger_\uparrow(i)\rangle
+\langle \psi_\uparrow(0) ,{\mathcal C}^\dagger(i)\psi_\downarrow(i)\rangle\nonumber\\
&&+ \langle {\mathcal C}(0)\psi^\dagger_\downarrow(0), \psi^\dagger_\uparrow(i)\rangle
+ \langle {\mathcal C}(0)\psi^\dagger_\downarrow(0) ,{\mathcal C}^\dagger(i)\psi_\downarrow(i)\rangle ,
\label{c2comf}
\end{eqnarray}
whose momentum transformation satisfies (\ref{cprop}). 
A similar result holds 
for the spin-down composite fermion
$\Psi_\downarrow(i)=R_B^{-1/2}\psi_\downarrow(i) 
+ R_B^{-3/2}{\mathcal C}(i)\psi^\dagger_\uparrow(i)$. 
They can be represented in the effective Hamiltonian 
\begin{eqnarray}
\!\!{\mathcal H}^F_{\rm eff} \!=\!\! \!\sum_{i,\sigma=\uparrow\downarrow}(\ell^3)Z^{-1}_F\Psi_{\sigma}^\dagger(i)\! \Big[\!-\!\nabla^2/(2M_F\ell^2) \!-\!\mu_F\Big]\!\Psi_{\sigma}(i).
\label{ceHamif}
\end{eqnarray}
Here $\mu_F =-M_F/2$ is the 
 chemical  potential and 
            $Z_F=g/M_F$ the wave-function 
renormalization.
Following the renormalization
(\ref{reno}) of fermion fields,
we renormalize composite fermion field $\Psi_{\uparrow,\downarrow}
\Rightarrow (Z_F)^{-1/2} \Psi_{\uparrow,\downarrow}$, which behaves as a quasi-particle in Eq.~(\ref{ceHamif}), analogously to the composite boson (\ref{eHamib}). The negatively charged ($e$) three-fermion state 
is a negatively charged ($2e$) paring field 
${\mathcal C}(i)=\psi_\downarrow (i)\psi_\uparrow(i)$
of two fermions combining with a hole $\psi^\dagger_\downarrow(i)$. 
These negatively charged ($e$)
composite fermions $\Psi_{\uparrow\downarrow}(i)$ 
are composed of three-fermion states 
${\mathcal C}\psi^\dagger_{\uparrow}$ or 
${\mathcal C}\psi^\dagger_{\downarrow}$
and a fermion $\psi_{\uparrow}$ or  $\psi_{\downarrow}$. 
Similarly, positively charged ($-e$)
composite fermions $\Psi^\dagger_{\uparrow}(i)$ or 
$\Psi^\dagger_{\downarrow}(i)$  
are composed by three-fermion states 
${\mathcal C}^\dagger\psi_{\uparrow}$ or
${\mathcal C}^\dagger\psi_{\downarrow}$ 
combined with a
hole $\psi^\dagger_{\uparrow}$ or
$\psi^\dagger_{\downarrow}$. 
Suppose that two constituent fermions  
$\psi_\downarrow(k_1)$ and $\psi_\uparrow(k_2)$ of the paring field 
${\mathcal C}(q)$, one constituent hole $\psi^\dagger_\uparrow(k_3)$ are around the Fermi surface, 
$k_1\approx k_2\approx k_3\approx k_F$, then the paring field 
${\mathcal C}(q)$ for
$q=k_2-k_1\ll k_F$, and three-fermion state 
$p=k_1-k_2+k_3\approx k_3\approx k_F$ is around the Fermi surface. As a result, the composite fermions $\Psi_{\uparrow\downarrow}$ live around the Fermi surface as well. Suppose that the three-fermion state 
$p=k_1-k_2+k_3\approx k_3\ll k_F$, the composite fermion propagator (\ref{cpropf}) reduces to its continuum version $S_{\rm Fermion}(p)\sim {(p^2+M_F^2)^{-1}}$.

As discussed below Eq.~(\ref{scalar0_c}), in the regime of small coupling $g$, the form factor $Z_B$ of composite boson
vanishes and the bosonic bound 
state pole dissolves into two fermionic constituent cut. Analogously, the form factor $Z_F$ of composite fermion vanishes and the 
three-fermion bound state pole dissolves into three fermionic constituent cut \cite{doubling}, i.e., three unpaired fermions. The smooth cross-over transition takes place.

The same results (\ref{csll})--(\ref{ceHamif}) are obtained for the  plasmon field 
${\mathcal P}(i)=\psi^\dagger_\downarrow (i)\psi_\uparrow(i) $ combined
with another fermion or hole, and the associated composite fermion
\begin{eqnarray}
\Psi^{\mathcal P}_\uparrow(i)&=&R_B^{-1/2}\psi_\uparrow(i) + R_B^{-3/2}{\cal P}(i)\psi_\downarrow(i)\nonumber\\
&\Rightarrow & g^{1/4}\psi_\uparrow(i) + g^{3/4}{\cal P}(i)\psi_\downarrow(i), 
\label{comf}
\end{eqnarray}
whose two-point Green function,
\begin{eqnarray}
&&\langle \Psi^{\cal P}_\uparrow(0) ,\Psi^{\cal P\dagger}_\uparrow(i)\rangle
=\langle \psi_\uparrow(0), \psi^\dagger_\uparrow(i)\rangle
+\langle \psi_\uparrow(0) ,{\cal P}^\dagger\psi^\dagger_\downarrow(i)\rangle\nonumber\\
&&+ \langle {\cal P}\psi_\downarrow(0), \psi^\dagger_\uparrow(i)\rangle
+ \langle {\cal P}\psi_\downarrow(0) ,{\cal P}^\dagger\psi^\dagger_\downarrow(i)\rangle.
\label{2comf}
\end{eqnarray}
The same is for $\Psi^{\cal P}_\downarrow(i)=R_B^{-1/2}\psi_\downarrow(i) + R_B^{-3/2}{\cal P}(i)\psi_\uparrow(i)$ 
the spin-down field. 
They can be represented in the effective Hamiltonian (\ref{ceHamif}) 
with $\Psi_{\sigma}(i)\rightarrow\Psi_{\sigma}^{\cal P}(i)$,
\comment{as
\begin{eqnarray}
\beta{\mathcal H}^{\cal P}_{\rm eff} &=& \beta\sum_{i,\sigma=\uparrow\downarrow}(\ell^3)Z^{-1}_F\Psi_{\sigma}^{\cal P\dagger}(i) \Big[-\nabla^2/(2M_F\ell^2) -\mu_F\Big]\Psi^{\cal P}_{\sigma}(i).
\label{eHamif}
\end{eqnarray}
The  chemical potential is $\mu_F =-M_F/2$, and the 
form factor $Z_F=g/M_F$.
}
following the renormalization (\ref{reno}) of fermion fields, 
and renormalization $\Psi^{\cal P}_{\uparrow,\downarrow}\Rightarrow (Z_F)^{-1/2} \Psi^{\cal P}_{\uparrow,\downarrow}$. The charged three-fermion states ${\cal P}\psi_{\uparrow\downarrow}$ or ${\cal P}^\dagger\psi^\dagger_{\uparrow\downarrow}$ are composed of a fermion or a hole combined with 
a neutral plasmon field ${\cal P}(i)=\psi^\dagger_\downarrow (i)\psi_\uparrow(i)$ or ${\cal P}^\dagger(i)=\psi^\dagger_\uparrow (i)\psi_\downarrow(i)$ 
of a fermion and a hole. 
The composite fermions $\Psi^{\cal P}_{\uparrow,\downarrow}(i)$ 
are composed of a three-fermion state 
${\cal P}\psi_{\uparrow,\downarrow}$ in combination with a fermion $\psi_{\uparrow}$ or $\psi_{\downarrow}$. 
The same thing is true for
its charge-conjugate state. 
Suppose that constituent fermion 
$\psi_\downarrow(k_1)$ and hole $\psi^\dagger_\downarrow(k_2)$,
another constituent fermion $\psi_\uparrow(k_3)$ 
are all around the Fermi surface, 
$k_1\approx k_2\approx k_3\approx k_F$, and the plasmon field
$q=k_2-k_1\ll k_F$ and composite fermion $p=k_1-k_2+k_3\approx k_3\approx k_F$ is around the Fermi surface as well.
The three-fermion states in Eqs.~(\ref{ccomf}) and (\ref{comf}) are related, ${\mathcal C}(i)\psi^\dagger_\downarrow(i)=-{\cal P}(i) \psi_\downarrow (i)$.
\comment{can be written as 
\begin{eqnarray}
${\mathcal C}(i)\psi^\dagger_\downarrow(i)
&=&\psi_\downarrow (i)\psi_\uparrow(i)\psi^\dagger_\downarrow(i)
=-\psi^\dagger_\downarrow(i)\psi_\uparrow(i) \psi_\downarrow (i)\nonumber\\
&=&-{\cal P}(i) \psi_\downarrow (i)$.
\label{caeq}
\end{eqnarray}
}
This implies that the three-fermion states  
${\mathcal C}(i)\psi^\dagger_\downarrow(i)$  and  ${\cal P}(i)\psi_\downarrow(i)$ are the same up to a definite 
phase factor $e^{i\pi}$. Thus the composite fermions $\Psi_\sigma(i)$ (\ref{ccomf}) and 
$\Psi^{\cal P}_\sigma(i)$ (\ref{comf}) are indistinguishable up to a definite phase factor. 

\comment{
All composite fermions are of the Dirac type, due to the interaction (\ref{sca}). 
For stronger couplings $1/ak_F > (1/ak_F)_{qc}$, 
Eq.~(\ref{csml}) should be extended by more complex composite fermions
$
$
$
S_{ML}(i)\hspace{-3pt}\equiv\hspace{-3pt}\langle \psi_\uparrow(0) ,
{\mathcal C}^\dagger(i)\psi_\downarrow(i)\rangle 
$
$
+~\langle \psi_\uparrow(0), {\mathcal S}_{\Uparrow}(i) \psi^\dagger_\uparrow(i)\rangle,
$
where $ {\mathcal S}_{\Uparrow}(i)=\psi^\dagger_\uparrow(i) \psi_\uparrow(i)$
brings in a spin-vector field.
}

\vskip0.1cm
\section
{\bf Conclusion and Remarks}
\hskip0.01cm
We present some discussions of the critical temperature (\ref{26h}), effective 
Hamiltonians (\ref{eHamib}) and (\ref{ceHamif}).
(i) In the regime $T^*> T >T_c$, there is a mixed liquid of composite bosons and fermions with 
the pseudogap $M_{F,B}(T)$, which is expected to dissolve to normal unpaired Fermi gas at the crossover temperature $T^*$. 
The non-interacting composite 
bosons and fermions are either charged or neutral, obeying 
Bose-Einstein or Fermi-Dirac distribution.
They behave as
superfluids up to a relatively high crossover temperature $T^*$. 
(ii) In the regime $T_c > T$, 
composite bosons develop a nonzero $\langle\mathcal C\rangle$ and 
undergo BEC, the ground state contains the BEC-dressing fermions
$\Psi^{\rm B}_\uparrow$ and $\Psi^{\rm B}_\downarrow$, which are 
composite fermions of Eqs.~(\ref{ccomf}-\ref{ceHamif}) 
by substituting composite boson ${\mathcal C}$ 
for its expectation value $\langle\mathcal C\rangle$.
These massive fermionic quasi-particles 
pair tightly to new bosonic quasi-particles 
$\Phi_{\rm B}=\Psi^{\rm B}_\uparrow\Psi^{\rm B}_\downarrow$ or 
$\Psi^{{\rm B}\dagger}_\uparrow\Psi^{\rm B}_\downarrow$,
which undergo a BEC condensate $\langle \Phi_{\rm B}\rangle\not=0$ to minimize the ground-state energy. This may be the origin of high-$T_c$ superconductivity and a similar form of composite superfluidity.
\comment{no scenes to referee A: ``In both cases, if the Coulomb repulsion between electrons could be compensated
by ``phonons'' in an analogous way to 
a Feshbach resonance or in $\Phi_{\rm B}$ by a composite-fermion pair state, this would result in superconductivity and superfluity at high temperature $T_c\propto {\mathcal O}(\epsilon_F)$.''
The coherent supercurrents consist of composite fermions dressed by composite
bosons.}  
If there is a smooth crossover transition from BSC to BEC, these features, although discussed for $1/ak_F> 0$, 
are expected to be also
true in $1/ak_F\ll 0$ with much smaller scale $M_{F,B}(T)$.
Due to the presence of composite fermions in addition to composite bosons, we expect a further suppression of the low-energy spectral weight for single-particle excitations and the material follows harder equation of state. Its observable consequences include a further $T$-dependent suppression of heat capacity and gap-like dispersion in the density-of-states and spin susceptibility. The measurements of entropy 
per site \cite{optical}, vortex-number \cite{Yao2016}, and shift in dipole oscillation frequency \cite{Roy2016} can be probes into
composite-boson and -fermion spectra, and pairing mechanism.

It is known that the limit 
$1/ak_F\ll 0$ produces an IR-stable fixed point, and
its scaling domain is described by an effective 
Hamiltonian of BCS physics with the gap scale $\Delta_0=\Delta(T_c)$ in $T\sim T_c\lesssim T^*$. This is
analogous to the IR-stable fixed point and scaling domain
of an effective Lagrangian of Standard Model (SM) with the
electroweak scale in 
particle physics \cite{xueJHEP2016,SHKN}. 

The unitarity limit $1/ak_F\rightarrow 0^\pm$ 
representing a scale invariant point \cite{UVscaling} was
formulated in a renormalization group framework \cite{NS2007}, implying
a UV-stable fixed point of large coupling, where
the running coupling $g$ 
approaches $g_{\rm UV}$, as the energy scale 
becomes large. 
In 
the scaling domain of this UV-fixed point 
$1/ak_F\rightarrow 0^\pm$ and $T\rightarrow T_c^u$, an effective 
Hamiltonian of composite bosons and fermions is realized with characteristic scale 
\begin{equation}
M_{B,F}(T)=\left(\frac{T-T^u_c}{T^u_c}\right)^{\nu/2} \frac{(2d)^{1/2}}{(3\pi^2)^{1/d}}k_F,\quad T\gtrsim T^u_c
\label{bfm}
\end{equation}
with a critical exponent  $\nu=1$ \cite{KS}. 
This shows the 
behavior of the
correlation length $\xi\propto M^{-1}_{B,F}$,
and characterizes 
the size of composite particles via the
 wave function renormalization factor 
$Z_{B,F}\propto M^{-1}_{B,F}$. 
This domain should be better explored experimentally.   
The analogy was discussed in 
particle physics 
with
anticipations of the UV scaling domain at TeV scales and effective 
Lagrangian of composite particles made by SM elementary fermions 
\cite{xueJHEP2017}. 

\section{\bf Methods}
\vskip0.1cm
\noindent{\bf Strong-coupling limit and expansion.}
\hskip0.1cm
To calculate the expansion in
strong-coupling 
limit, we relabel
$\beta \ell^d\rightarrow  \beta $ and $2m\ell^2 \rightarrow 2m$, 
so the lattice spacing $\ell$ is set equal to unity, and rescales 
$\psi_\sigma(i)\rightarrow (\beta g)^{1/4}\psi_\sigma(i)$ and
$\psi^\dagger_\sigma(i)\rightarrow (\beta g)^{1/4}\psi^\dagger_\sigma(i)$, 
so that the Hamiltonian (\ref{Hami}) can be written as $\beta{\mathcal H} 
= \sum_i\left[h {\cal H}_0(i)+{\cal H}_{\rm int}(i)\right]$, 
where the hopping parameter $h\equiv \beta/(\beta g)^{1/2}
$ and 
\begin{eqnarray}
\!\!\!\!\!\!\! {\cal H}_0(i)\! &=&\!
\sum_{\sigma=\uparrow,\downarrow}\psi^\dagger_\sigma(i) (-\nabla^2/2m -\mu)\psi_\sigma(i),\label{sHami0}\\
\!\!\!\!\!\!\!{\cal H}_{\rm int}(i)\!&\equiv&\! -
\psi^\dagger_\uparrow(i) \psi^\dagger_\downarrow(i) \psi_\downarrow(i)\psi_\uparrow(i),
\label{sHami}
\end{eqnarray}
and the partition function with expectation values
\begin{eqnarray}
\!\!\!\!{\mathcal Z} &=& \Pi_{i,\sigma}
\int d\psi_\sigma(i)d\psi^\dagger_\sigma(i) \exp(-\beta{\mathcal H}),
\label{part}\\
\!\!\!\!\langle\cdot\cdot\cdot\rangle \!&=&\!{\mathcal Z}^{-1} \Pi_{i,\sigma}
\int d\psi_\sigma(i)d\psi^\dagger_\sigma(i)(\cdot\cdot\cdot)\exp(-\beta{\mathcal H}). 
\end{eqnarray}
Fermion fields $\psi_\uparrow$ and $\psi_\downarrow$ are one-component 
Grassmann variables with $\psi_\sigma(i) \psi_{\sigma'}(j) =- \psi_{\sigma'}(j)\psi_\sigma(i)$ and integrals
$\!\!\!\int\! d\psi_\sigma(i) \psi_{\sigma'}(j)
=\delta_{\sigma,\sigma'}\delta_{ij},$ 
$\int\! d\psi^\dagger_\sigma(i) \psi^\dagger_{\sigma'}(j)
=\delta_{\sigma,\sigma'}\delta_{ij}.$
All others vanish. 

In the 
limit $h\rightarrow 0$ 
for $g\rightarrow\infty$ and finite $T$, 
the kinetic terms (\ref{sHami0}) are neglected, 
and the partition function (\ref{part})
has a nonzero strong-coupling limit
\begin{eqnarray}
\!\!\Pi_i\!\!\int_{i\downarrow}\int_{i\uparrow}\exp(-{\cal H}_{\rm int})\!=\!
-\Pi_i\!\!\int_{i\downarrow}\psi_\downarrow(i)^\dagger
\psi_\downarrow(i)=(1)^{\mathcal N} , 
\end{eqnarray}
where ${\mathcal N}$ is the total number of lattice sites, 
$\int_{i\uparrow}\equiv
\int[d\psi_\uparrow^{\dagger}(i)d\psi_\uparrow(i)]$ and 
$\int_{i\downarrow}\equiv
\int[d\psi_\downarrow^\dagger(i)d\psi_\downarrow(i)]$.
The strong-coupling expansion
can now be performed in powers of the hopping parameter $h$. 

\vskip0.1cm
\noindent{\bf Green functions of composite particles.}
\hskip0.01cm
The leading strong-coupling
approximation to Green function (\ref{bosonpc}) is
$G(i) = \delta ^{(d)} (i)/ \beta g$.
The first correction is obtained
by using the one-site partition function $Z(i)$ and the integral 
\begin{eqnarray}
&&\!\!\!\!\!\!\!\!\!\!\langle \psi_\uparrow\psi_\downarrow\rangle
\!\equiv  \!{1\over Z(i)}\int_{i\downarrow}\int_{i\uparrow}
\psi_\uparrow(i)\psi_\downarrow(i)e^{-h{\cal H}_0(i)-\beta{\cal H}_{\rm int}(i)}\label{p2}
\\
&&\!\!={h^2}\sum^\ave_{\hll}\psi_\uparrow(i;\hll)
\sum^\ave_{\hll'}\psi_\downarrow(i;\hll')
\!\approx \!  {h^2}
\sum^\ave_{\hll}\psi_\uparrow(i;\hll)\psi_\downarrow(i;\hll),\nonumber 
\end{eqnarray}
~\\[-1.3em]
where the non-trivial result needs
$\psi^\dagger_{\uparrow,\downarrow}(i)$ 
fields in the hopping expansion of $e^{-h{\cal H}_0(i)}$, and
$
\sum^\ave_{\hll}\psi_\sigma(i;\hll)
\equiv\sum_{\hll}\left[\psi_\sigma(i+\hll)+
\psi_\sigma(i-\hll)\right].
$
In Eq.~(\ref{bosonpc}), integrating over fields $\psi_{\uparrow,\downarrow}(i)$ 
at the site ``$i$'', the first corrected version reads:
\begin{eqnarray}
\vspace{-2em}
\hspace{-2.5em}G(i)&= &{\delta^{(d)}(i)\over \beta g}
+{1\over \beta g}\left({\beta\over 2m
}\right)^2~\sum^\ave_{\hll}G^{\rm nb}(i;\hll),
\vspace{-3.5em}
\label{re4}
\end{eqnarray}
~\\[-.8em]
where $\delta^{(d)} (i)$ is a spatial $\delta$-function and 
$G^{\rm nb}(i\pm\hll)$ is the Green function (\ref{bosonpc}) without integration over fields $\psi_{\sigma}$ at the neighbor site $i$. Note that 
the nontrivial contributions come only from kinetic hopping terms 
$\propto (h/2m)^2 =(1/\beta g)(\beta/2m)^2$, 
excluding the chemical potential term 
$\mu \psi^\dagger_\sigma(i)\psi_\sigma(i)$ in the Hamiltonian 
(\ref{sHami0}). 

Replacing $G^{\rm nb}(i\pm\hll)$ by $G(i\pm\hll)$ converts 
Eq.~(\ref{re4}) into
a recursion relation for $G(i)$, which actually takes into account of 
 high-hopping 
corrections in a strong-coupling expansion. Going to momentum space 
we approximately obtain $G({q})={1\over \beta g}+{2\over \beta g}\left({\beta\over 2m
}\right)^2G({ q})\sum_{\hll}\cos ( q{\hll})$, 
solved by
\begin{eqnarray}
G({ q})&=& {\left[{2m/(\beta\ell)
}\right]^{2}\over 4\ell^{-2}\sum_{\hll}\sin^2(q\hll/2)+M_B^2 }.
\label{rep4'}
\end{eqnarray}
Here we have resumed the original lattice spacing $\ell$ by
setting back  
$\beta\rightarrow \beta\ell^3$ and $2m\rightarrow 2m\ell^2$.

By the analogy to (\ref{p2})--(\ref{rep4'}), we calculate Green functions (\ref{csll}-\ref{csmm}) for composite fermions and obtain three recursion relations 
\begin{eqnarray}
\hspace{-.5cm}S_{LL}(p)\!\!&=&\!\!{1\over \beta g}\left({\beta\over 2m
}\right)^3\Big[2\sum_{\hll} \cos (p{\hll})\Big] S_{ML}(p),
\label{crep1}\\
\hspace{-.5cm}S_{ML}(p)\!\!&=&\!\!{1\over \beta g}
+{1\over \beta g}\left({\beta\over 2m
}\right)\Big[2\sum_{\hll} \cos ({p\hll})\Big] S_{LL}(p),
\label{crep2}\\
\hspace{-.5cm}S_{MM}(p)\!\!&=&\!\!{1\over \beta g}\left({\beta\over 2m
}\right)\Big[2\sum_{\hll} \cos ({p\hll})\Big] S^{\dagger}_{ML}(p).
\label{crep3}
\end{eqnarray}
We solve these recursion relations 
and obtain
\begin{eqnarray}
\!\!\!\!\!S_{ML}(p)&=&\frac{(1/\beta g)}{1-(1/\beta g)^2(\beta/2m)^4\Big[2\sum_{\hll} \cos ({p\hll})\Big]^2},
\label{crep2'}
\end{eqnarray}
$S_{LL}(p)$ and $S_{MM}(p)$. 
Define for the propagator of the composite fermion
$S_{\rm Fermion}(p)$ the quantity $g S(p)$,
\begin{eqnarray}
S(p)&=&R_B^{-1}S_{LL}(p) + 2R_B^{-2}S_{ML}(p)+R_B^{-3}S_{MM}(p)\nonumber\\
&=& {2\over 4\ell^{-2}\sum_{\hll}\sin^2(p\hll/2)+M_F^2 },
\label{cprop}
\end{eqnarray}
where $R_B$ and $M_F^2=M_B^2$ follow Eq.~(\ref{scalar}). For some more details, please see Ref.~\cite{kx2017}.

\comment{
\vskip0.1cm
\noindent{\bf Author contributions}\\
Both authors conceived the project, discussed the results 
and prepared the manuscript, S.-S.~Xue performed calculations.
\vskip0.1cm
\noindent{\bf Additional information}\\
Correspondence and requests for materials should be addressed to S.-S.~Xue.
\vskip0.1cm
\noindent{\bf Competing financial interests}\\
The authors declare no competing financial interests.
}

\end{document}